\documentclass[prx,nofootinbib,twocolumn]{revtex4-1}
\usepackage{amsmath,srcltx}
\usepackage{mathtools, amssymb} 
\usepackage{color}  
\usepackage{graphicx}
\usepackage{float}
\def\a{\alpha}
\def\b{\beta}
\def\d{\delta}
\def\ve{\varepsilon}
\def\f{\phi}
\def\F{\Phi}

\def\g{\gamma}
\def\h{\eta}

\def\j{\psi}

\def\l{\lambda}
\def\m{\mu}
\def\n{\nu}

\def\r{\rho}

\def\s{\sigma}

\def\x{\xi}

\def\D{\Delta}

\def\G{\Gamma}

\def\Om{\Omega}

\def\S{\Sigma}

\def\cf{{\cal F}}

\def\co{{\cal O}}

\def\cs{{\cal S}}

\def\sl#1{\rlap{\hbox{$\mskip 1 mu /$}}#1}
\def\Sl#1{\rlap{\hbox{$\mskip 3 mu /$}}#1}

\newcommand{\be}{\begin{equation}}
\newcommand{\ee}{\end{equation}}
\newcommand{\ba}{\begin{eqnarray}}
\newcommand{\ea}{\end{eqnarray}}

\newcommand{\ov}{\overline}
\newcommand{\uv}{\underline}

\newcommand{\wh}{\widehat}

\newcommand{\aand}{\;\;\;\mbox{and}\;\;\;}
\newcommand{\pa}{\partial}

\def\I{\leavevmode\hbox{\small1\kern-3.8pt\normalsize1}}

\DeclareMathOperator{\e}{e}

\newcommand{\pari}{\stackrel{{P}}\longrightarrow}

\newcommand{\mc}{\mathcal}

\begin{document}
\title{The quantum scale invariance in graphene-like quantum electrodynamics}

\author{O.M. Del Cima} \email{oswaldo.delcima@ufv.br} 

\author{D.H.T. Franco} \email{daniel.franco@ufv.br}

\author{L.S. Lima} \email{lazaro.lima@ufv.br}

\author{E.S. Miranda} \email{emerson.s.miranda@ufv.br} 

\affiliation{Universidade Federal de Vi\c cosa (UFV),\\
Departamento de F\'\i sica - Campus Universit\'ario,\\
Avenida Peter Henry Rolfs s/n - 36570-900 - Vi\c cosa - MG - Brazil.}
\affiliation{Ibitipoca Institute of Physics (IbitiPhys),\\
36140-000 - Concei\c c\~ao do Ibitipoca - MG - Brazil.}


\begin{abstract}
The ultraviolet and infrared finiteness of a parity-even massless planar quantum electrodynamics mimics the scale invariance in graphene.      
\end{abstract}


\maketitle

{\it In honor of the 80th birthday of Prof. Olivier Piguet}
\\

\emph{Introduction.--} The quantum electrodynamics in three space-time dimensions (QED$_3$) has called attention since the groundbreaking works by Schonfeld, Jackiw, Templeton and Deser \cite{deser-jackiw-templeton-schonfeld} owing to the viability of taking planar quantum electrodynamics models as theoretical foundation for quasiplanar condensed matter phenomena, such as high-$T_{\rm c}$ superconductors \cite{high-Tc}, quantum Hall effect \cite{quantum-hall-effect}, topological insulators \cite{topological-insulators}, topological superconductors \cite{topological-superconductors} and graphene \cite{graphene}. Since then, planar quantum electrodynamics models have been investigated in many physical arrangements, namely, small (perturbative) and large (non perturbative) gauge transformations, abelian and nonabelian gauge groups, fermions families, odd and even under parity, compact space-times, space-times with boundaries, curved space-times, discrete (lattice) space-times, external fields and finite temperatures. Besides, from the perturbative renormalization point of view, three dimensional space-time quantum field theories exhibit remarkable features, like superrenormalizability and finiteness.  

Preliminary results at 1-loop order \cite{one-loop} and later on at 2-loops \cite{two-loops}, especially in Chern-Simons models \cite{deser-jackiw-templeton-schonfeld}, have shown how interesting perturbative finiteness should be. At all orders in perturbation theory, pure nonabelian Chern-Simons model in the Landau gauge manifests ultraviolet finiteness \cite{chern-simons-landau-gauge}. Although coupled to fermionic and bosonic matter fields, nonabelian Chern-Simons model in three space-time Riemannian manifolds still exhibits ultraviolet finiteness at all radiative order \cite{chern-simons-matter}. Beyond that, the massless $U(1)$ QED$_3$ exhibits ultraviolet and infrared perturbative finiteness, parity and infrared anomaly free at all orders \cite{masslessU1QED3}. Moreover, despite of some claims found out in the literature about that parity could spontaneously be broken, even perturbatively, in massless $U(1)$ QED$_3$, called parity anomaly, has been rejected by the consistent and correct use of dimension regularization \cite{rao-delbourgo}, Pauli-Villars regularization \cite{pimentel}, algebraic renormalization in the framework of Bogoliubov-Parasiuk-Hepp-Zimmermann-Lowenstein (BPHZL) subtraction method \cite{masslessU1QED3}, and through the Epstein-Glaser method \cite{epstein-glaser}. The quantum scale invariance in dimensional reduced to three dimensional space-time massless QED$_4$ models has been presented in \cite{dudal-mizher-pais}, and the gauge covariance of the massless fermion propagator was studied in quenched QED$_3$ \cite{quenchedQED3}. 

The pristine graphene, a monolayer of pure graphene \cite{graphene}, is a gapless bidimensional system behaving like a half-filling semimetal where the quasiparticles charge carriers are described by massless charged Dirac fermions. The electron-electron interactions in graphene \cite{electron-pairing} include electron-polarons \cite{electron-phonon} scattering processes \cite{polarons}, where the quasiparticle electron-polaron (or hole-polaron) is formed by a bound state of electron (or hole) and phonon \cite{landau}. The anomalous quantum Hall effect and fourfold degeneracy of Landau levels have been experimentally observed in \cite{landau-levels-exp} and theoretically described in \cite{masslessU1U1QED3}. Moreover, experimental data about the strong nonlinear optical response in graphene \cite{optical-response} points toward to quantum scale invariance and the absence of pseudochiral anomaly \cite{scale-invariance}, which are the both issues that shall be demonstrated throughout this letter for the model introduced in \cite{masslessU1U1QED3}.

\emph{The model.--} 
The parity-preserving $U_A(1)\times U_a(1)$ hybrid QED$_3$ \cite{masslessU1U1QED3} -- the pristine graphene-like quantum electrodynamics -- with the parity-even Lowenstein-Zimmermann mass term added, is given by:
\ba
\S^{(s-1)} &=& \int{d^3 x} \bigg\{-\frac{1}{4}F^{\m\n}F_{\m\n} -\frac{1}{4}f^{\m\n}f_{\m\n}  \nonumber \\
&+& \m \ve^{\m\a\n}A_\m\pa_\a a_\n +i {\ov\j_+} {\Sl D}\j_+ + i {\ov\j_-} {\Sl D}\j_-  \nonumber \\
&+& b \pa^\m A_\m+\frac{\a}{2}b^2+\ov{c}\square c + \pi \pa^\m a_\m+\frac{\b}{2}\pi^2+\ov{\x}\square \x \nonumber \\
&&\underbrace{-~ m(s-1){\ov\j_+}\j_+ +m(s-1){\ov\j_-}\j_-}_{\textrm{\small Lowenstein-Zimmermann mass term}}  \bigg\}~, \label{action} 
\ea
where ${\Sl D}\j_\pm \!\equiv\!(\sl\pa + ie\Sl{A} \pm ig\sl{a})\j_\pm$, $m$ and $\m$ are mass parameters with mass dimension $1$ and the coupling constants $e$ (electric charge) and $g$ (pseudochiral charge) are dimensionful with mass dimension $1/2$. The field strengths, $F_{\m\n}=\pa_\mu A_\nu - \pa_\n A_\m$ and $f_{\m\n}=\pa_\mu a_\nu - \pa_\n a_\m$, are related to the electromagnetic field ($A_\m$) and the pseudochiral gauge field ($a_\m$), respectively. The Dirac massless spinors $\j_+$ and $\j_-$ are two kinds of fermions where the $\pm$ subscripts are related to their pseudospin sign \cite{binegar,massiveU1U1QED3,masslessU1U1QED3}. However, due to the presence of massless fermions, $\j_+$ and $\j_-$, the momentum subtraction scheme modified by Lowenstein-Zimmermann (BPHZL) \cite{low} has to be adopted in order to deal with the infrared divergences that shall arise in the process of ultraviolet subtractions. Also, the fields $c$ and $\x$ are two kind of ghosts\footnote{It shall be pointed out that neither the ghosts ($c$ and $\x$) nor the antighosts ($\ov{c}$ and $\ov{\x}$) take part of any Feynman diagram since they are free quantum fields, thus they decouple.} and, $\ov{c}$ and $\ov{\x}$, the two antighosts, whereas $b$ and $\pi$ are the Lautrup-Nakanishi fields \cite{lautrup-nakanishi} playing the role of Lagrange multiplier fields for the gauge conditions. The adopted gamma matrices are $\g^\m=(\s_z,-i\s_x,i\s_y)$. Finally, the Lowenstein-Zimmermann parameter $s$ lies in the interval $0\le s\le1$ and has the same status of an additional subtraction variable (as the external momentum) in the BPHZL renormalization scheme, in such a way that the massless model \cite{masslessU1U1QED3} is recovered by taking $s=1$ at the end of calculations.

The action $\S^{(s-1)}$ (\ref{action}) is invariant under the Becchi-Rouet-Stora (BRS) transformations \cite{brs}:
\ba
&s\j_\pm=i(c \pm \x)\j_\pm ~,~ s\ov{\j}_\pm=-i(c \pm \x)\ov{\j}_\pm~;& \nonumber \\
&\displaystyle sA_\m=-\frac{1}{e}\pa_\m c ~,~ s c=0 ~;~ \displaystyle sa_\m=-\frac{1}{g}\pa_\m \x ~,~ s\x=0~;& \nonumber \\
&\displaystyle s\ov{c}=\frac{b}{e} ~,~ sb=0~;~ \displaystyle s\ov{\x}=\frac{\pi}{g} ~,~ s\pi=0~.&   \label{BRS}
\ea
However, in order to control at the quantum level the renormalization of the nonlinear BRS transformations, a parity-even antifields action ($\S_{\rm ext}$) has to be added to $\S^{(s-1)}$ (\ref{action}):
\ba
\S_{\rm ext} &=& \int{d^3x}\biggl\{i\ov{\Om}_+s\j_+ - i\ov{\Om}_-s\j_-  \nonumber \\
&-& s\ov{\j}_+\Om_+ +s\ov{\j}_-\Om_- \biggr\}~. \label{action_ext}
\ea
So, the complete BRS invariant tree-level action reads:
\begin{equation}
\G_0 \equiv \G^{(s-1)}_0 = \S^{(s-1)}+\S_{\rm ext}~.\label{completeaction}
\end{equation}
In time, it should be pointed out that in spite of the Faddeev-Popov ghosts be massless, which could result in serious infrared divergences through radiative corrections, since they are free fields there is no need to introduce Lowenstein-Zimmermann mass terms for them \cite{low}. Finally, the action $\G_0$ (\ref{completeaction}) is invariant under parity transformations ($P$), which upon the fields and antifields are given by:
\ba
\j_\pm  & \pari & \j_\pm^P=-i\g^1\j_\mp~,~~\ov\j_\pm \pari \ov\j_\pm^P=i\ov\j_\mp\g^1~,       \nonumber \\
\Om_\pm  & \pari & \Om_\pm^P=-i\g^1\j_\mp~,~~\ov\Om_\pm \pari \ov\Om_\pm^P=i\ov\Omega_\mp\g^1~,  \nonumber \\
A_\mu  & \pari & A_\mu^P=(A_0,-A_1,A_2)~,                                                    \nonumber \\
a_\mu  & \pari & a_\mu^P=(-a_0,a_1,-a_2)~,                                                   \nonumber \\
\phi       & \pari & \phi^P=\phi~,~~\phi=\{b, \bar{c}, c\}~,           \nonumber \\
\chi      & \pari &     \chi^P=-\chi~, ~~\chi=\{\pi, \bar{\xi}, \xi\}~. \label{parity_transformation}
\ea

By switching off the coupling constants ($e=g=0$) in $\G_0$ (\ref{completeaction}), the following propagators show up:
\begin{align}
&\D_{\pm \pm} = i\dfrac{{\sl k}\mp m(s-1)}{k^2-m^2(s-1)^2}~,~~~~~~~~~~~~~ \label{propk++--massless}             \\   
&\D^{\m\n}_{AA} = -i\bigg\{\dfrac{1}{k^2-\mu^2}\left(\h^{\m\n}-\frac{k^\m k^\n}{k^2}\right)+\frac{\a}{k^2}\dfrac{k^\m k^\n}{k^2}\bigg\}, ~ \label{propkAAmassless}         \\          
&\D^{\m\n}_{aa} = -i\bigg\{\dfrac{1}{k^2-\mu^2}\left(\h^{\m\n}-\frac{k^\m k^\n}{k^2}\right)+\frac{\b}{k^2}\dfrac{k^\m k^\n}{k^2}\bigg\},   ~ \label{propkaamassless}        \\
& \D_{Aa}^{\m\n} = \dfrac{\mu}{k^2(k^2-\mu^2)}\e^{\mu\a\n}k_\a ~,                                              \label{propkAamaslesss}               \\
&\D_{Ab}^\m = \D_{a \pi}^\m(k) = \dfrac{k^\m}{k^2}~,~     \label{propkAbmassless}                                                                   \\
&\D_{bb} = \D_{\pi\pi}(k) = 0~,                                                                                                                    \\
&\D_{\ov{c}c}= \D_{\ov{\x} \x}(k) = -\frac{i}{k^2}~.    \label{propkccmasless}       
\end{align}
Here it is opportune to mention that, from these propagators the spectrum content, {\it i.e.} degrees of freedom, spin, masses
and charges, as well as its consistency, namely the tree-level unitarity and causality, have been analyzed in \cite{masslessU1U1QED3}.

The ultraviolet (UV) and infrared (IR) dimensions of any fields, for instance $X$ and $Y$, are established by means of the UV and IR asymptotic behavior of their propagator $\D_{XY}(k)$, $d_{XY}$ and $r_{XY}$, respectively, then
\be
d_{XY}={\ov{\rm deg}}_{k}\D_{XY}(k) \aand r_{XY}={\uv{\rm deg}}_{k}\D_{XY}(k)~,
\ee
where ${\ov{\rm deg}}_{k}$ gives the asymptotic behavior of the propagator when $k, s\rightarrow \infty$, while ${\uv{\rm deg}}_{k}$ gives its asymptotic behavior when $k\rightarrow 0$ and $s\rightarrow 1$. The UV ($d$) and IR ($r$) 
dimensions of the fields, $X$ and $Y$, satisfy the following inequalities \cite{brs, piguet-sorella}:
\be
d_X + d_Y \geq 3 + d_{XY} \aand r_X + r_Y \leq 3 + r_{XY}~. \label{uv-ir}
\ee
Thereby, from the propagators (\ref{propk++--massless}) associated to the Dirac spinors, $\psi_+$ and $\psi_-$, together with the previous conditions (\ref{uv-ir}), we get
\ba
 && d_{\pm \pm}=-1~ \Rightarrow~ 2d_\pm\geq 2~   \Rightarrow~ d_\pm=1~;\label{d+}\\ 
 && r_{\pm\pm}=-1 \Rightarrow~ 2r_\pm \leq 2~    \Rightarrow~ r_\pm=1~.  \label{r+}   
\ea
Similarly, considering the propagators (\ref{propkAAmassless}), (\ref{propkaamassless}) and (\ref{propkAamaslesss}) related to the gauge fields, $A_\m$ and $a_\m$, and the constraints (\ref{uv-ir}), in the Landau gauge $\alpha=\beta=0$, we have
\ba
 &&d_{Aa}=-3~ \Rightarrow~ d_A + d_a\geq 0~; \label{dAa}\\
 &&d_{AA}=-2~ \Rightarrow~ 2d_A\geq 1~ \Rightarrow~ d_A=\frac{1}{2}~;\\
 &&d_{aa}=-2~ \Rightarrow~ 2d_a\geq 1~ \Rightarrow~ d_a=\frac{1}{2}~; \label{dAAaa}\\
 &&r_{Aa}=-1~ \Rightarrow~ r_A + r_a\leq 2~; \label{rAa} \\
 &&r_{AA}=0~ \Rightarrow~ 2r_A\leq 3~ \Rightarrow~ r_A=1~;\\
 &&r_{aa}=0~ \Rightarrow~ 2r_a\leq 3~ \Rightarrow~ r_a=1~. \label{rAAaa}
\ea
It is important to mention that, notwithstanding the mixed propagator (\ref{propkAamaslesss}) brings no on-shell degrees of freedom \cite{masslessU1U1QED3}, whenever it enters as an internal (off-shell) line of any 1-particle irreducible Feynman diagram, provided it exhibits non-negative UV degree of divergence ($\d\geq 0$), BPHZL ultraviolet and infrared subtractions have to be performed. The UV and IR dimensions of all fields, as well as their ghost number and Grassmann parity are summarized in Table \ref{table_1}.

\begin{table}[]
\begin{center}
\begin{tabular}{|c|c|c|c|c|c|c|c|c|c|c|c|c|c|c|}
\hline
         &$A_\mu$ &$a_\mu$ & $\j_+$ & $\j_-$ &$c$  &${\ov c}$ &      $b$    &  $\x$    & ${\ov\x}$ &     $ \pi$    &  $s$  & $s-1$  & $\Om_+$   &$\Om_-$  \\
\hline
$d$      &${1/2}$ &${1/2}$ &  1     &   1    & 0   &    1     &$\frac{3}{2}$&    0     &    1     &$\frac{3}{2}$   &   1   &   1    &      2    &     2   \\
\hline
$r$      &    1   &   1    &   1    &   1    &  0  &    1     &      1      &    0     &    1     &      1         &   0   &   1    &      2    &     2   \\
\hline
$\F\Pi$  &   0    &   0    &  0     &   0    & 1   &  $-1$    &     0       &    1     &   $-1$   &      0         &   0   &   0    &     $-1$  &   $-1$   \\
\hline
$GP$     &   0    &   0    &  1     &   1    & 1   &    1     &     0       &    1     &   1      &      0         &   0   &   0    &       0   &    0     \\
\hline
\end{tabular}
\end{center}
\caption[]{UV dimension ($d$), IR dimension $(r)$, ghost number ($\F\Pi$) and Grassmann parity ($GP$).}\label{table_1}
\end{table}

In order to represent the symmetries in a functional way, {\it e.g.} the BRS symmetry and the rigid symmetries, we begin with the Slavnov-Taylor operator ($\cs$) acting on an arbitrary functional ($\mc{F}$):
\ba
\cs(\cf)&=&\int{d^3 x} \biggl\{-{1\over e}{\pa}^\mu c {\d\cf\over\d A^\mu} + {b\over e} {\d\cf\over\d {\ov c}} 
- {1\over g}{\pa}^\mu \x {\d\cf\over\d a^\mu} + {\pi\over g} {\d\cf\over\d {\ov \x}} \nonumber \\
&+& {\d\cf\over\d \ov\Om_+}{\d\cf\over\d \j_+} - {\d\cf\over\d \Om_+}{\d\cf\over\d \ov\j_+} \nonumber \\
&-& {\d\cf\over\d \ov\Om_-}{\d\cf\over\d \j_-} + {\d\cf\over\d \Om_-}{\d\cf\over\d \ov\j_-} \biggl\}~,\label{slavnov}
\ea
such that if $\mc{F}=\G_0$, it leads to 
\be
\cs(\G_0)=0~.\label{slavnovident}
\ee
The corresponding linearized Slavnov-Taylor operator ($\cs_\cf$) associated to (\ref{slavnov}) reads:
\begin{widetext}
\ba
\cs_\cf &=&\int{d^3 x} \biggl\{-{1\over e}{\pa}^\mu c {\d\over\d A^\mu} + {b\over e} {\d\over\d {\ov c}} 
-{1\over g}{\pa}^\mu \x {\d\!\over\d a^\mu} + {\pi\over g} {\d \over\d {\ov \x}} 
+{\d\cf\over\d \ov\Om_+}{\d\over\d \j_+} + {\d\cf\over\d \j_+}{\d\over\d \ov\Om_+}                                      
- {\d\cf\over\d \Om_+}{\d\over\d \ov\j_+} - {\d\cf\over\d\ov\j_+}{\d\over\d \Om_+} \nonumber  \\
&-&{\d\cf\over\d \ov\Om_-}{\d\over\d \j_-} - {\d\cf\over\d \j_-}{\d\over\d \ov\Om_-} + {\d\cf\over\d \Om_-}{\d\over\d \ov\j_-} + {\d\cf\over\d\ov\j_-}{\d\over\d \Om_-} \biggl\}~,\label{slavnovlin}
\ea
\end{widetext}
where from (\ref{slavnov}) and (\ref{slavnovlin}) it follows that the next nilpotency identities holds:
\ba
&&\cs_\cf\cs(\cf)=0~,~~\forall\cf~,\label{nilpot1} \\
&&\cs_\cf\cs_\cf=0~~{\mbox{if}}~~\cs(\cf)=0~. \label{nilpot3}
\ea
In particular, thanks to the Slavnov-Taylor identity (\ref{slavnovident}), the linearized Slavnov-Taylor operator $\cs_{\G_0}$ is nilpotent, {\it i.e.} $(\cs_{\G_0})^2=0$. In addition to, the operator $\cs_{\G_0}$ acting upon the fields and the antifields are given by:
\ba
&&\cs_{\G_0}\f=s\f~,~~\f=\{\j_\pm,\ov\j_\pm,A_\m,a_\m, c,{\ov c},b,\pi,{\ov \x}, \x\}~,\nonumber\\
&&\cs_{\G_0}\Om_+=-\frac{\d\G_0}{\d\ov\j_+} ~,~~\cs_{\G_0}\ov\Om_+=\frac{\d\G_0}{\d\j_+}~,\nonumber\\
&&\cs_{\G_0}\Om_-=\frac{\d\G_0}{\d\ov\j_-}~,~~~~\cs_{\G_0}\ov\Om_-=-\frac{\d\G_0}{\d\j_-}~. \label{operation1}
\ea
Additionally, the tree-level action $\G_0$ (\ref{completeaction}) satisfies the ghost equations, the antighost equations and the gauge conditions displayed below:
\ba
-i\frac{\d \G_0}{\d c} &=&  i\Box\ov{c}+\ov{\Om}_+\j_+-\ov{\Om}_-\j_-+\ov{\j}_+\Om_+-\ov{\j}_-\Om_-~, \nonumber\\  
-i\frac{\d \G_0}{\d \x} &=& i\Box\ov{\x}+\ov{\Om}_+\j_++\ov{\Om}_-\j_-+\ov{\j}_+\Om_++\ov{\j}_-\Om_-~, \nonumber \\  
\frac{\d \G_0}{\d \ov{c}} &=& \Box c   ~,~~ \frac{\d \G_0}{\d \ov{\x}}= \Box \x~,  \nonumber \\ 
\frac{\d \G_0}{\d b} &=& \pa^\m A_\m +\a b~,~~ \frac{\d \G_0}{\d \pi}=\pa^\m a_\m +\b \pi~. \label{ghost_equation4}
\ea
To end the analysis about the symmetry content of the model described by the classical action $\G_0$ (\ref{completeaction}), it should be stressed that it is also invariant under two rigid symmetries stemming from the gauge symmetry $U_A(1)\times U_a(1)$, that express indeed the conservation of electric charge and pseudochiral charge:
\be
W_{\rm rigid}^{(e)} \G_0=0 \aand W_{\rm rigid}^{(g)} \G_0=0~, \label{rigidcond}
\ee
with the Ward operators $W^{(e)}_{\rm rigid}$ and $W^{(g)}_{\rm rigid}$ been given as:
\ba
W_{\rm rigid}^{(e)}&=&\int{d^3 x}\biggl\{\j_+{\d\over\d \j_+}-\ov\j_+{\d\over\d \ov\j_+}+\Om_+{\d\over\d \Om_+}-\ov\Om_+{\d\over\d \ov\Om_+}  \nonumber\\
&+&\j_-{\d\over\d \j_-} - \ov\j_-{\d\over\d \ov\j_-} + \Om_-{\d\over\d \Om_-} - \ov\Om_-{\d\over\d \ov\Om_-}\biggr\}~,\label{wrigid_e} \\
W_{\rm rigid}^{(g)}&=&\int{d^3 x}\biggl\{\j_+{\d\over\d \j_+}-\ov\j_+{\d\over\d \ov\j_+}+\Om_+{\d\over\d \Om_+}-\ov\Om_+{\d\over\d \ov\Om_+}  \nonumber\\
&-& \j_-{\d\over\d \j_-} + \ov\j_-{\d\over\d \ov\j_-} - \Om_-{\d\over\d \Om_-} + \ov\Om_-{\d\over\d \ov\Om_-}\biggr\}~.\label{wrigid_g}
\ea

\emph{Searching for anomalies.--}
The stability of a classical action under perturbations, namely the multiplicative renormalizability, does not enable a priori its extension to the quantum level once there is no guarantee about the absence of any gauge anomaly, {\it i.e.} electromagnetic and pseudochiral anomalies. The so-called parity anomaly, which is often cited in the literature as a typical  anomaly even perturbatively in three space-time dimensions, does not show up in this model as shall be proved in the sequence.

The quantum vertex functional ($\G$) can be expanded as a formal power series in $\hbar$, in such a way that at zeroth order it coincides with the classical action $\G_0$ (\ref{completeaction}):
\be
\G \equiv \G^{(s-1)}=\G_0 + {\co}(\hbar)~,\label{vertex}
\ee
and ought to satisfy the same constraints (\ref{ghost_equation4})--(\ref{rigidcond}) as the tree-level action. A classical symmetry which is broken at the quantum level is called an anomaly. Although some anomalies could not compromise necessarily the spectrum consistency of the model, once they simply express that the corresponding symmetries do not hold anymore at the quantum level, there is a class of anomalies that violates the $S$-matrix unitarity, spoiling, as a consequence, the spectrum consistency, they are the consistent anomalies. 

A consistent anomaly might stem from the Slavnov-Taylor identity if it is broken at the quantum level. In other words, if the Slavnov-Taylor identity holds up to order $n-1$ in $\hbar$, meaning that  the broken shows up at $\hbar^{n}$ order, $\cs(\G)|_{s=1}=\mathcal{O}(\hbar^{n})$, so according to the quantum action principle {\cite{piguet-sorella}}:
\be
\cs(\G)=\hbar^n\D \cdot \G|_{s=1} = \hbar^n\D + {\co}(\hbar^{n+1})~, \label{slavnovbreak}
\ee
where $\D\equiv \D^{(s-1)}|_{s=1}$ is a local integrated functional in the fields and antifields, with ghost number $1$ and UV and IR dimensions bounded by, $d\le 7/2$ and $r\geq 3$, respectively (see Table \ref{table_1}). Accordingly, the Slavnov-Taylor breaking (\ref{slavnovbreak}) would be an anomaly, provided it could not be reabsorbed into the quantum action as a noninvariant counterterm. 

Bearing in mind the equation $\cs_{\G}=\cs_{\G_0} + {\co}(\hbar)$ derived from (\ref{slavnovlin}) and (\ref{vertex}), together with the nilpotency ({\ref{nilpot1}) and the Slavnov-Taylor quantum breaking ({\ref{slavnovbreak}) identities, it yields the Wess-Zumino consistency condition for the quantum breaking $\D$: 
\be
\cs_{\G_0}\D=0~.\label{breakcond1}
\ee
In addition to that, besides the Wess-Zumino consistency condition (\ref{breakcond1}), by taking into consideration the Slavnov-Taylor identity (\ref{slavnovident}), the ghost, antighost and gauge equations (\ref{ghost_equation4}), as well as the rigid conditions (\ref{rigidcond}), the Slavnov-Taylor quantum breaking (\ref{slavnovbreak}) has to satisfy the following constraints:
\ba
&&  {\d\D\over\d b}={\d\D\over\d\ov c}=0~,    ~\int d^3x \frac{\d\D}{\d c}=0    ~,~~ W_{\rm rigid}^{(e)}\D=0~,  \nonumber \\
&&  {\d\D\over\d \pi}={\d\D\over\d\ov{\x}}=0~,~\int d^3x \frac{\d\D}{\d \x}=0   ~,~~ W_{\rm rigid}^{(g)}\D=0~.  \label{breakcond5}
\ea
Furthermore, the conditions (\ref{breakcond5}) imply that the local integrated functional $\D$ (\ref{slavnovbreak}) does not dependent explicitly on the fields $b$, $\pi$, $\ov{c}$, $\ov{\x}$, $c$ and $\x$, while for the two latter ghost fields their  functional dependence is of the form $\partial_\m c$ and $\partial_\m \x$. At this moment, it should pointed out that, concerning the invariance of $\D$ under the rigid symmetries $W_{\rm rigid}^{(e)}$ and $W_{\rm rigid}^{(g)}$ as displayed in (\ref{breakcond5}), some comments shall be made owing to the fact that the symmetry group $U_A(1)\times U_a(1)$ is a non-semisimple Lie group, thus in principle the associated rigid symmetry might be anomalous. Nonetheless, as long as the both abelian factors are not spontaneously broken, and the electric charge ($e$) and the pseudochiral charge ($g$) are conserved, $W_{\rm rigid}^{(e)} \G_0=0$ and $W_{\rm rigid}^{(g)} \G_0=0$, as a result the conditions, $W_{\rm rigid}^{(e)}\D=0$ and $W_{\rm rigid}^{(g)}\D=0$, are still fulfilled \cite{stora,kraus}. 

Reminding the Wess-Zumino consistency condition (\ref{breakcond1}), it constitutes a cohomology problem in the sector of ghost number one. Actually, the general solution to this cohomology problem can always be written as a sum of a trivial cocycle $\cs_{\G_0}{\wh\D}^{(0)}$, where ${\wh\D}^{(0)}$ has ghost number zero, and of a nontrivial term having ghost number one, ${\wh\D}^{(1)}$, which lies in the cohomology of the linearized Slavnov-Taylor operator $\cs_{\G_0}$ (\ref{slavnovlin}), then:
\be
\D = {\wh\D}^{(1)} + \cs_{\G_0}{\wh\D}^{(0)}~,\label{breaksplit}
\ee
such that, as previously mentioned, the Slavnov-Taylor quantum breaking $\D$ (\ref{breaksplit}) has to satisfy the conditions (\ref{breakcond1}) and (\ref{breakcond5}). It should be remarked that the local field integrated polynomial belonging to the functional $\wh \D^{(0)}$, can be added order by order to the vertex functional $\G$, that is $\cs_{\G_0}(\G-\hbar^n{\wh\D}^{(0)}) = \hbar^n{\wh\D}^{(1)} + {\co}(\hbar^{n+1} \D)$, as a noninvariant integrated local counterterm, $-\hbar^n{\wh\D}^{(0)}$. Accordingly, there will be an anomaly at any order if $\wh{\D}^{(1)}\neq 0$, otherwise the quantum breaking might be absorbed order by order restoring therefore the Slavnov-Taylor identity.   

In order to pursue our searching for possible anomalies, we draw attention to UV ($\d$) and IR ($\r$) power-counting of this model for any 1-particle irreducible Feynman diagram \cite{parityconservation}:
\ba
\!\!\!\!\!\!\!\!\d &=&3-\sum_{f}d_f N_f-\sum_{b}d_b N_b-\frac{N_e+N_g}{2}-N_{Aa}~, \nonumber \\
\!\!\!\!\!\!\!\!\r &=&3-\sum_{f}r_f N_f-\sum_{b}\frac{3r_b}{2} N_b+\frac{N_e+N_g}{2}-N_{Aa}~,\label{power_counting_massless}
\ea
where $N_f$ and $N_b$ are the numbers of external lines of fermions and bosons, respectively, $N_e$ is the power in $e$, and $N_g$ the power in $g$, whereas $N_{Aa}$ is the number of internal lines associated to the mixed propagator $\D^{\m\n}_{Aa}$. 
Seeing that the radiative corrections are at least of 1-loop order, so they are at least of order two in the coupling constants, namely $e^2$, $g^2$ or $eg$, as a consequence, thanks to the power-counting (\ref{power_counting_massless}), the quantum breaking $\D$ (\ref{breaksplit}) is bounded indeed by the  UV and IR dimensions, $d\le 5/2$ and $r\geq 4$.

Recalling the constraints (\ref{breakcond5}) the Slavnov-Taylor quantum breaking $\D$ (\ref{breaksplit}) has to fulfill, we conclude that:
\be
\D = \int{d^3 x}~\left\{ {\cal K}^{(0)}_\mu\pa^\m c+ {\cal X}^{(0)}_\m\pa^\m \x \right\}~, \label{delta_1}
\ee
where ${\cal K}^{(0)}_\mu$ and ${\cal X}^{(0)}_\m$ are rank one tensors with ghost number 0, also they are UV and IR bounded by $d\leq 3/2$ and $r\geq 3$, respectively. Beyond that, as already discussed and proved in \cite{parityconservation}, the BPHZL subtraction procedure does not break parity for the model presented here, implying for that reason that the breaking $\D$ (\ref{delta_1}) shall be parity-even. Therefore, we conclude that $\mathcal{K}_\m^{(0)}$ and $\mathcal{X}_\m^{(0)}$, which can be expressed by $\mathcal{K}_\m^{(0)}=\sum_{i} v_{k,i}\mathcal{V}^i_\m$ and $\mathcal{X}_\m^{(0)}=\sum_{i} v_{x,i}\Pi^i_\m$, are such that $\mathcal{V}^i_\m$ are vectors, whereas $\Pi^i_\m$ are pseudovectors, with $v_{k,i}$ and $v_{x,i}$ being fixed scalar coefficients. Let us now write down all possible candidates for vectors $\mathcal{V}^i_\m$ and pseudovectors $\Pi^i_\m$:
\ba
&&  {\cal V}^{1}_\m = A_\m A^\n A_\n ~,~ {\cal V}^{2}_\m = A_\m a^\n a_\n ~,~ {\cal V}^{3}_\m = A_\n a^\n a_\m ~, 
\nonumber \\
&&  {\Pi}^{1}_\m = a_\m a^\n a_\n ~,~ {\Pi}^{2}_\m = a_\m A^\n A_\n ~,~ {\Pi}^{3}_\m = a_\n A^\n A_\m ~, 
\ea
bringing $\D$ (\ref{delta_1}) to be written as follows:
\ba
\D&=&\int d^3x  \bigg\{[v_{k,1}A_\m A^\n A_\n+v_{k,2}A_\m a^\n a_\n \nonumber\\ 
&+& v_{k,3}A_\n a^\n a_\m] \partial^\m c + [v_{x,1}a_\m a^\n a_\n \nonumber\\ 
&+& v_{x,2}a_\m A^\n A_\n+v_{x,3}a_\n A^\n A_\m] \partial^\m \xi\bigg\}~.\label{anomaly_even_1}
\ea
Besides, we verify that the expression above, for the possible anomaly $\D$ (\ref{anomaly_even_1}), can be rewritten as: 
\be
\D=\cs_{\G_0} \big(\l_1 {\wh\D}^{(0)}_1 + \l_2 {\wh\D}^{(0)}_2 + \l_3 {\wh\D}^{(0)}_3 + \l_4 {\wh\D}^{(0)}_4\big)~,\label{anomaly_even_2}
\ee
such that the local integrated monomials ${\wh\D}^{(0)}_i$ ($i=1,\dots,4$) are given by:
\ba
\nonumber
&&\wh\D^{(0)}_1 = \int d^3x (A_\m A^\m)^2~,~ \wh\D^{(0)}_2= \int d^3x (a_\m a^\m)^2~, \nonumber\\
&&\wh\D^{(0)}_3 = \int d^3x A_\m A^\m a_\n a^\n~, \nonumber\\ 
&&\wh\D^{(0)}_4 = \int d^3x A_\m A^\n a_\n a^\m~, \label{delts}
\ea
with 
\ba
&& v_{k,1}=-\frac{4}{e}\l_1 ~,~~ v_{k,2}=-\frac{2}{e}\l_3 ~,~~ v_{k,3}=-\frac{2}{e}\l_4 ~,~ \nonumber \\ 
&& v_{x,1}=-\frac{4}{g}\l_2 ~,~~ v_{x,2}=-\frac{2}{g}\l_3 ~,~~ v_{x,3}=-\frac{2}{g}\l_4 ~.
\ea
Finally, we have demonstrated that as far as the presence of gauge anomaly is concerned, the monomials ${\wh\D}^{(0)}_i$ (\ref{delts}) can be incorporated order by order into the quantum action as noninvariant counterterms, for instance at $\hbar^n$-order:
\ba 
\cs_{\G_0}(\G-\hbar^n \D) &\equiv & \cs_{\G_0}\left(\G-\hbar^n \sum_{i=1}^4 \l_i {\wh\D}^{(0)}_i\right) \nonumber \\ 
&=& 0\hbar^n + {\co}(\hbar^{n+1})~, \label{a_final}
\ea
which finishes the proof about the absence of gauge anomaly, {\it i.e.} $\{{\wh\D}^{(1)}\}\equiv \varnothing$. It is opportune to mention that, although the absence of gauge anomaly, the model might be suffered from infrared anomaly stemming from the massless fermions, however, none of the monomials ${\wh\D}^{(0)}_i$ ($i=1,\dots,4$) violates the infrared condition $r\geq 4$. In summary, it has been proved that the parity-preserving $U_A(1)\times U_a(1)$ hybrid QED$_3$ \cite{masslessU1U1QED3} is free from any anomalies (${\wh\D}^{(1)}=0$), arising from either discrete or continuous symmetries, in all radiative order. Though to complete the quantum level analysis, it remains to check the multiplicative renormalizability of the model, which will be presented in the sequence.

\emph{Searching for counterterms.--}
The multiplicative renormalizability, the so-called stability condition\footnote{The stability condition deals also with the issue of looking for the most general classical action compatible with all symmetries of the proposed model.}, is accomplished whenever perturbative quantum corrections produce only local counterterms corresponding to the renormalization of parameters that are already present in the tree-level action, so those counterterms can be reabsorbed order by order through redefinitions of the initial physical quantities, {\it i.e.} coupling constants, masses, and fields. That being the case, in order to check if the tree-level action $\G_0$ (\ref{completeaction}) is stable perturbatively, it is deformed by an integrated local polynomials in the fields and antifields (counterterm) $\G^c\equiv\G^{c(s-1)}$, {\it i.e.} $\G^\ve=\G_0+\ve \G^c$, with $\ve$ being an infinitesimal parameter, while the counterterm action $\G^c$ possesses the same quantum numbers as the tree-level action $\G_0$ (\ref{completeaction}) when $s=1$. Keeping in mind the above proof about the absence of any kind of anomaly, namely gauge anomaly, rigid anomaly, infrared anomaly, or parity anomaly, once it indicates that all classical continuous and discrete symmetries are assured at the quantum level, the perturbed action $\G^\ve$ fulfills all the classical action ($\G_0$) constraints (\ref{slavnovident}), (\ref{ghost_equation4}) and (\ref{rigidcond}). Consequently, the counterterm $\G^c$ is subjected to the constraints below:   
\ba
&&\cs_{\G_0}\G^c=0~,~~ \G^c \pari  \G^c~; \nonumber\\
&& W_{\rm rigid}^{(e)}\G^c=0 ~,~~ W_{\rm rigid}^{(g)}\G^c=0~; \nonumber\\
&&\frac{\d\G^c}{\d b}=\frac{\d\G^c}{\d c}=\frac{\d\G^c}{\d \ov{c}}=
\frac{\d\G^c}{\d \pi}=\frac{\d\G^c}{\d \x}=\frac{\d\G^c}{\d \ov{\x}}=0~; \nonumber\\
&&\frac{\d \G^c}{\d\Om_+}=\frac{\d \G^c}{\d\ov{\Om}_+}=\frac{\d \G^c}{\d\Om_-}=\frac{\d \G^c}{\d\ov{\Om}_-}=0~. \label{stabcond4} 
\ea 
Additionally, based on the BPHZL subtraction scheme framework, which is parity invariant for this model \cite{parityconservation}, the most general local field polynomial ($\G^c$) satisfying the conditions (\ref{stabcond4}), and with UV and IR dimensions bounded by $d\leq 7/2$ and $r\geq 3$, respectively, reads:      
\ba
\G^c &=& \int  {d^3 x}  \bigg\{\upsilon_1 i{\ov\j}_+{\Sl D}\j_+ +  \upsilon_1 i{\ov\j}_-{\Sl D}\j_- \nonumber \\ 
&+&\upsilon_2 F^{\m\n}F_{\m\n} + \upsilon_3 f^{\m\n}f_{\m\n} + \upsilon_4  \epsilon^{\m\a\n}A_\m \pa_\a a_\n\bigg\}~, \label{finalcount}
\ea
where the arbitrary parameters $\upsilon_i$ ($i=1,\dots,4$) have to be fixed later by normalization conditions. Due to the superrenormalizability of the model, manifested by the ultraviolet and infrared coupling-constant-dependent power-counting (\ref{power_counting_massless}), together with the fact that counterterms arise from loop corrections, they are at least of order two in the coupling constants, namely, $e^2$, $g^2$ or $eg$. In fact, because of that, the effective UV and IR dimensions of the counterterm $\G^c$ (\ref{finalcount}) have to be kept within the bounds $d\leq 5/2$ and $r\geq 4$. The latter condition implies the vanishing of all arbitrary parameters $\upsilon_i=0$ ($i=1,\dots,4$), meaning that the typical ambiguities owing to any renormalization procedure do not show up in the present model, {\it i.e.} $\{\G^c\}\equiv \varnothing$. At the end of the day, this result, namely the absence of any counterterm ($\G^c=0$), combined with the previous one about the absence of any kind of anomaly (${\wh\D}^{(1)}=0$), finishes the proof upon the ultraviolet and infrared (full) perturbative finiteness -- vanishing all $\beta$-functions and all anomalous dimensions -- as well as the absence of parity and infrared anomaly. Ultimately, the issue of quantum scale invariance and absence of pseudochiral anomaly exhibited by the parity-even $U_A(1)\times U_a(1)$ hybrid QED$_3$ \cite{masslessU1U1QED3} looks to agree with experimental data \cite{optical-response,scale-invariance}.

\emph{Conclusion.--}
As a final conclusion, we have shown that the pristine graphene-like quantum electrodynamics \cite{masslessU1U1QED3} is free from any gauge and global, infrared, and parity anomaly at all radiative orders. Likewise, the model presents vanishing gauge coupling ($e$ and $g$) constants and Chern-Simons mass ($\m$) parameter $\b$-functions, $\b_e=\b_g=\b_\m=0$, and vanishing anomalous dimensions of all the fields, $\g=0$. It should be stressed that the proof introduced here is independent of any specific regularization procedure since the BRS algebraic renormalization method \cite{brs,piguet-sorella,stora} jointly with the BPHZL subtraction scheme \cite{low} is supported by general theorems of perturbative quantum field theory. A final matter deals with the proof of the quantum consistency of the model displayed above, and some remarkable behaviors of monolayer pure graphene -- anomalous quantum Hall effect and fourfold degeneracy of Landau levels \cite{landau-levels-exp}, absence of pseudochiral anomaly and quantum scale invariance \cite{optical-response,scale-invariance} -- the model embodies. The two former effects have been analyzed in \cite{masslessU1U1QED3}, whereas the two latter, the absence of pseudochiral anomaly and quantum scale invariance, were theoretically revealed in this letter.

\emph{Acknowledgements.--}
The authors dedicate this work to the 80th birthday of Prof. Olivier Piguet. CAPES-Brazil is acknowledged for invaluable financial help.

\end{document}